\documentclass[preprint]{revtex4}
\usepackage{graphicx}
\usepackage{amsmath}
\usepackage{amssymb}
\usepackage{color}
\usepackage[normalem]{ulem}
\usepackage{epsfig}
\usepackage{amssymb,amsfonts,amsmath,color}
\usepackage{array}
\usepackage{makecell}

\makeatletter
\def\thickhline{%
  \noalign{\ifnum0=`}\fi\hrule \@height \thickarrayrulewidth \futurelet
   \reserved@a\@xthickhline}
\def\@xthickhline{\ifx\reserved@a\thickhline
               \vskip\doublerulesep
               \vskip-\thickarrayrulewidth
             \fi
      \ifnum0=`{\fi}}
\makeatother

\newlength{\thickarrayrulewidth}
\newcommand{\stkout}[1]{\ifmmode\text{\sout{\ensuremath{#1}}}\else\sout{#1}\fi}

\begin{document}

\title{Evolutionary dynamics in populations with fluctuating size}

\author{Immanuel Meyer and Nadav M. Shnerb}

\affiliation{Department of Physics, Bar-Ilan University,
Ramat-Gan IL52900, Israel.}

\begin{abstract}
\noindent Temporal environmental variations are ubiquitous in nature, yet most of the theoretical works in population genetics and evolution assume fixed environment. Here we analyze the effect of variations in carrying capacity on the fate of a mutant type. We consider a two-state Moran model, where selection intensity at equilibrium may differ (in amplitude and in sign) from selection  during periods of sharp growth and sharp decline.  Using Kimura's diffusion approximation we present simple formulae for effective population size and effective selection, and use it to calculate the chance of ultimate fixation, the time to fixation and the time to absorption (either fixation or loss). Our analysis shows perfect agreement with numerical solutions for neutral, beneficial and deleterious mutant. The contributions of different processes to the mean and the variance of abundance variations are additive and commutative. As a result, when selection intensity $s$ is weak such that ${\cal O}(s^2)$ terms are negligible, periodic or stochastic environmental variations yield identical results.
\end{abstract}

\maketitle

\section{Introduction}

Evolution takes place in a temporally fluctuating environment, and the interplay between the deterministic effect of selection and stochastic effects associated with environmental variations poses a major theoretical challenge. Even if the macro properties of the environment are fixed over time, local fluctuations affect the reproductive success of individuals in an uncorrelated manner (genetic drift, demographic stochasticity) and generate, for a population of size $N$, an ${\cal O}(\sqrt{N})$ noise. When the macro-environment varies, it affects coherently the demographic rates of entire populations and generates ${\cal O}(N)$ fluctuations~\cite{lande2003stochastic}.

Traditionally, the theory of population genetics and evolution was focused on the interference of  selection (with intensity $s$) with demographic stochasticity (drift)~\cite{parsons2010some}, assuming fixed birth and death rates. Effects of environmental variations were considered only rarely~\cite{takahata1975effect,takahata1979genetic}. Recent empirical studies have documented periodic and stochastic coherent variations in relative fitness~\cite{bergland2014genomic,bell2010fluctuating,messer2016can} as well as variations  in the birth and death rates~\cite{caceres1997temporal,hoekstra2001strength,leigh2007neutral,hekstra2012contingency,kalyuzhny2014niche,kalyuzhny2014temporal,chisholm2014temporal}. These findings triggered a renewed interest in the effect of macro-environmental variations on evolutionary dynamics~\cite{mustonen2008molecular,huerta2008population,ashcroft2014fixation,cvijovic2015fate,hidalgo2017species,danino2018fixation,meyer2018noise,marrec2019resist}.

The simplest and most important scenario in evolution involves zero-sum competition between two  haploid types (very similar model describes the dynamics of a two-allele, one locus system for diploid population with additive effect on fitness). The three properties that govern evolutionary dynamics are $\Pi(x)$, the chance that a mutant type (with abundance $n$ and frequency $x=n/N$) will reach ultimate fixation, $T_A(x)$,  the time to absorption (either fixation or loss) and $T_F(x)$, the time to fixation. $\Pi(x)$  plays a determinant role in the evolutionary dynamics as it controls the long-term adaptation of populations and the rate of accumulation of neutral substitutions (molecular clock)~\cite{ewens2012mathematical}. $T_A(x)$ sets the timescale for coexistence and controls the crossover from  successive-fixation to clonal interference dynamics~\cite{desai2007speed}, and $T_F(x)$ governs the adaptation process and the speed of evolution~\cite{danino2018environmental}.

 For the sake of concreteness, one may imagine  a wild type and a mutant population competing for a single resource (say, food), where the relative fitness of the mutant type, reflecting its ability to consume or to reach the food, is $s$. $s$ is positive for a beneficial mutant, negative for a deleterious mutant and $s=0$ for a neutral mutant. Under fixed environmental conditions, the absolute value of $sN$ sets the strength of selection. When $|sN| < 1$ (weak selection) demographic fluctuations (drift) dominate and the dynamics is effectively neutral, when $|sN|>1$ selective forces dominate  (strong selection). For fixed $N$ and $s$  the solutions for these quantities are known for many years~\cite{crow1970introduction,ewens2012mathematical}.

 Here we consider the dynamics of a mutant population/allele, when the carrying capacity fluctuates in time due to seasonal fluctuations (such as  food shortages during winter) or stochastic disturbances (droughts, floods). In that case the carrying capacity decreases when the environment deteriorates and increases when the environmental conditions improve. Moreover, the selective parameters during periods of  growth or  decline may differ (in amplitude and in sign) from their equilibrium values. For example, larger seeds have larger quantities of metabolic reserves than smaller seeds while smaller seeds can be produced in larger quantities. Therefore, large seeds have a better chance of establishment during periods of increasing stresses whereas small seeds  have better chance to colonize a new suitable habitat during fast expansion~\cite{smith1974optimal}. To account for that, we made a distinction between $s$, the selection coefficient at equilibrium, and $s_g$ and $s_d$, the selection coefficient during growth or decline, correspondingly. 
 
 The relationships between these scenarios and the results obtained for fixed $N$ and fixed $s$~\cite{crow1970introduction,ewens2012mathematical} are not obvious. In particular the system may jump between $N$ values that correspond to weak selection and $N$ values that correspond to strong selection, and a beneficial mutant at equilibrium ($s>0$) may become effectively deleterious if $s_g$ and/or $s_d$ are negative and vice versa.

To address variations in $N$ and $s$, some  authors~\cite{engen2009fixation,uecker2011fixation} implemented Haldane's~\cite{haldane1927mathematical} branching process approximation. This method is limited to calculation of $\Pi$ (not $T_A$ or $T_F$), to the regime $x \ll 1$  and   to  beneficial mutations only.  Here we present a general solution which is based on careful implementation of Kimura's diffusion approximation. We calculated the deterministic change and the stochastic variance of $x$ during periods of growth or  decline, and add these quantities (with the correct weights that depend on the typical timescale between events, $\tau$) to the diffusion and to the convection terms in an appropriate Backward Kolmogorov  equation (BKE). This equation has the functional form of a static environment BKE, so one can implement the known result, replacing $s$ by an effective selection parameter  $s_{eff}$ and $N$ by an effective population size $N_{eff}$. Our main results,  Eqs. (\ref{eq1new}) and (\ref{eq1newsel}), provide $s_{eff}$ and $N_{eff}$ in terms of the process parameters.

This paper is organized as follows. In the next section we provide the details of the models to be considered, and emphasize the distinction between local and global competition and between stochastic and periodic variations.  In the third section we explain our analytic approach and derive our main results for the case of stochastic variations with local competition. Section \ref{sec4}  clarifies the conditions under which the usage of the diffusion approximation is valid, and explains why in this parameter regime Eqs. (\ref{eq1new}) and (\ref{eq1newsel}) are applicable to  other scenarios including global competition and periodic fluctuations. In the discussion section we clarify the relationships between our results and the recent works of Wienand et al~\cite{wienand2017evolution,wienand2018eco}, discuss the limitations of our methodology and its possible extensions.

\section{Model systems}

The size of a natural populations is usually determined by competition for a limiting resource (water, sunlight, food and so on). Resource density variations may reflect stochastic or periodic environmental changes, such as seasonality, global or local temperature variations, interspecific competition and predation pressure. In our model we distinguish between equilibrium dynamics (fixed population size) and short periods during which the carrying capacity varies.

We consider two types of zero-sum equilibrium dynamics: local and global.
\begin{itemize}
  \item The local dynamics corresponds to the case where a random encounter between individuals may involve a fight for a piece of food, a mate  or  a territory. To model that, two individuals are picked at random for a ``duel", the loser dies and the winner produces a single offspring. If the mutant frequency is $x$, the chance of a duel between a mutant and a wild-type is $2x(1-x)$. The chance of the mutant to win the duel is defined to be $1/2 + s/4$, so $s$ reflects the intensity of selection. When $s=0$ the equilibrium dynamics is neutral.

  \item Global dynamics best illustrates the competition in a forest, say, where adult tree dies at random and the gap is recruited by a single seed or seedling. If the local seed bank reflects the composition of the whole forest (long distance dispersal), the chance of the mutant type to capture the gap depends on both its abundance and its fitness.   In our global model this chance is, $$\frac{x e^s}{1-x+xe^s},$$
  where the fitness factor $e^s$ reflects an excess productivity of seeds or the excess chance of germination per seed.
\end{itemize}
With this parametrization, both models yield, to the leading order in $s$, the logistic behavior~\cite{meyer2018noise}
\begin{equation}
{\dot x} = sx(1-x).
\end{equation}
As explained below, this  allows us to implement the same formulas for the effective population size and the effective selection in both cases.

\begin{figure}[h]
	\centering{
		\includegraphics[width=8cm]{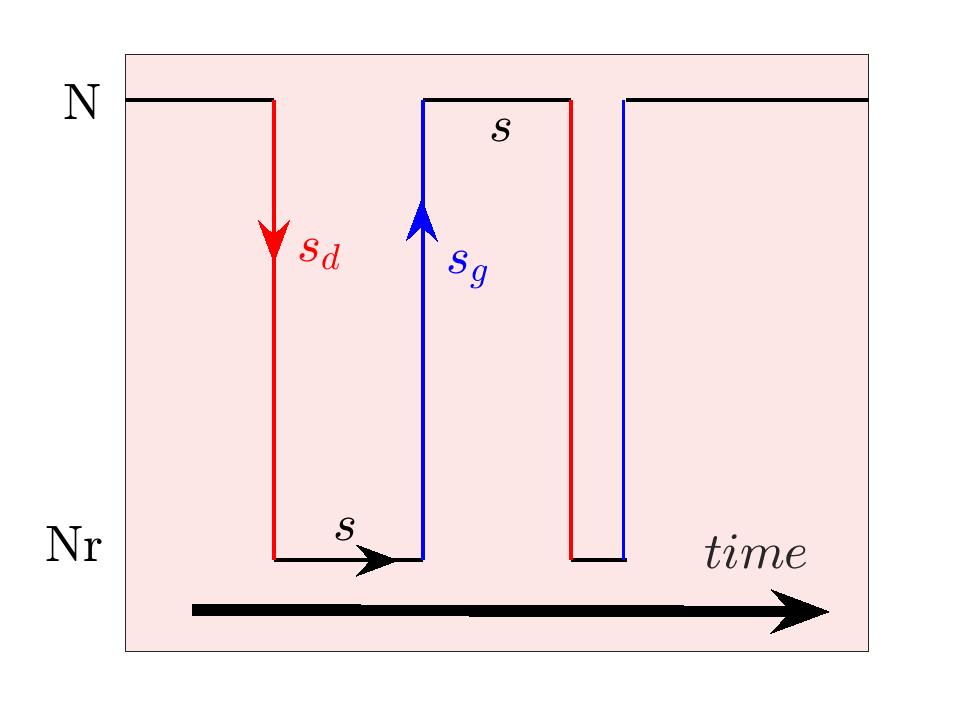}}
	\caption{Illustration of population size dynamics in varying environment. Under high resource density the total number of individuals is $N$, while during periods of shortage the total number is $rN$. Transitions between the two states correspond to periods  of sharp population  growth or sharp decline, here taken to be instantaneous. If the dynamics is periodic, the time between two consecutive jumps (between sharp growth to sharp decline or vise versa) is $\tau$. If the dynamics is stochastic (as in this cartoon) the sojourn time is picked from an exponential distribution with mean $\tau$. Selection parameter at equilibrium is $s$,  $s_g$ is the selection parameter during periods of sharp growth and $s_d$ is the selection parameter during sharp decline.      \label{fig0}}
\end{figure}

We model resource variations that affect the total carrying capacity through a simple two-state dynamics. When the resource density declines, the total population size decreases  from $N$ to $rN$ (without loss of generality we assume $r<1$), while an increase in the amount of available resource is followed by population growth from $rN$ back to $N$ (see Figure \ref{fig0}). In our model, carrying capacity variations are instantaneous; the limitations of this approximation are clarified in the  discussion section.

The persistence time of the environment is $\tau$ and the dynamics may be periodic or stochastic:
\begin{itemize}
  \item In periodically varying environment (seasonal variations) $\tau$ is the duration between two successive switches.
  \item In randomly fluctuating environment, the time between two successive switches is drawn from an exponential distribution with mean $\tau$.
\end{itemize}
At equilibrium, competition takes place in a series of birth-death events, and time is incremented by $1/N(t)$ after each of these events, so $\tau$ is measured in units of a generation (one generation $=$ $N$ elementary birth-death events).

During periods of sharp growth $(1-r)N$ slots open up, and the number of new recruit by the mutant strain is picked at random from  $B_{N(1-r)}[x+s_g x(1-x)]$, a binomial distribution with $N(1-r)$ trials where the chance to win each trial is,
 \begin{equation}
 \frac{x e^{s_g}}{x e^{s_g}+(1-x)} \approx x + s_g x (1-x).
 \end{equation}
 Accordingly, in a  period of sharp growth on average  $x \to  x + s_g x(1-x)(1-r)$ and the leading contribution to the variance is $Var(x) = x(1-x)(1-r)/N$.

During a period of sharp decline, each individual survives with a certain probability that may depend on its phenotype. To model that, we assumed the number of mutant survivors to be picked from $B_{xN}[r(1+s_d)]$ and  the number of wild type survivors is drawn from  $B_{(1-x)N}[r]$  (of course the condition $s_d \leq (1-r)/r$ must be imposed). If $s_d \ll 1$, in a period of sharp decline $x \to  x + s_d x(1-x)$ and $Var(x) = x(1-x)(1-r)/(Nr)$~\cite{wahl2001probability}.

\section{Effective population size and effective selection}

Under purely demographic stochasticity, when population size is fixed at $N$ and the mutant strain has fixed log-fitness $s$ and frequency $x=n/N$, the chance of fixation $\Pi(x)$ is known to satisfy~\cite{crow1970introduction,ewens2012mathematical},
\begin{equation} \label{eq0}
\frac{1}{2N^2} \Pi''(x) + \frac{s}{2N} \Pi'(x)=0,
\end{equation}
with the boundary conditions $\Pi(0) = 0$ and $\Pi(1) =1$. Similarly, the time to absorption $T_A$ satisfies,
\begin{equation} \label{eq0t}
\frac{1}{2N^2} T_A''(x) + \frac{s}{2N} T_A'(x)=-\frac{1}{Nx(1-x)},
\end{equation}
with $T_A(0) = T_A(1) = 0$.

The solution for $\Pi(x)$  is a known formula,
\begin{equation} \label{eq2new}
\Pi(x) = \frac{1-e^{-Nsx}}{1-e^{-Ns}}.
\end{equation}
  An important parameter is the strength of selection $sN$, which is the ratio between selection intensity and the strength of the drift (the variance of $x$ variations per generation $1/N$). As mentioned above, when $|sN| \ll 1$ (weak selection) the process is  effectively neutral, the sign of $s$ is irrelevant  and $\Pi(x) \approx x$. If $|sN| \gg 1$ (strong selection) the chance of a single deleterious mutant ($x=1/N$) to reach fixation decays exponentially with $sN$ while for a single beneficial mutant ($s>0$) $\Pi(1/N) \approx s$.

Now let us implement the diffusion approximation in varying environment. Specifically we would like to consider  \emph{stochastic} environmental variations with \emph{local} competition at equilibrium.
In the next section we explain why the outcome is applicable to the other cases (global competition, periodic variations)

In our model the carrying capacity flips instantaneously between its two allowed values, $N$ and $rN$, so during each elementary time-step either the system jumps ($N \leftrightarrow rN$)  or a duel between two individuals takes place.  Accordingly,  $J^+ = 1/N\tau$ and $J^- = 1/Nr\tau$ are the chances, per elementary competition step (duel), that the environment flips to the other state.

When a duel takes place, it will be an intraspecific duel with probability $1-2x(1-x)$ and an interspecific duel with probability $2x(1-x)$. Intraspecific duels has no effect on the number of mutant individuals (either a wild type replaces a wild type or a mutant replaces a mutant). After an interspecific duel, the number of mutant grows by one with probability $1/2+s/4$ or decreases by one with probability $1/2-s/4$.

When the environmental conditions suddenly improve, $rN \to N$, the overall population increases by $N(1-r)$ individuals. If the mutant and the wild type have equal fitness during a period of sharp expansion, the chance of the mutant species to win each open slot is proportional to its relative frequency $x$. Therefore, the overall demographic gain of the mutant is picked at random from $B_{N(1-r)}[x]$, the binomial distribution for with $N(1-r)$ trials and success probability $x$. Similarly, if the $s_g$ is the selection parameter during growth,  the share of the mutant species is taken to be  $B_{N(1-r)}[x+s_g x(1-x)]$. All in all, the chance of the mutant type to grow from $n$ individuals  to $m$ individuals  during a sharp growth  is,
 \begin{equation}
P^{growth}_{n \to m} = f_{N(1-r),x+s_g x(1-x)}(m-n) \qquad m \ge n,
\end{equation}
where  $f_{b,p}(a)$ is the probability mass function of the binomial distribution, i.e., the chance to pick $a$ successes from $b$ trials with success probability $p$.

During  periods of sharp decline $N \to rN$. If $s_d=0$, the chance of each individual (both mutant and wild type) to survive is $r$. If $s_d \neq 0$, wild type individual survives with probability $r$ and a mutant individual survived with probability $r(1+s_d)$.  Accordingly, if the number of mutant individuals before decline is $n$, the number after decline is picked from $B_{n}[r(1+s_d)]$, and correspondingly the number of wild type individuals after the decline is $B_{N-n}[r]$. Therefore,
 \begin{equation}
 P^{decline}_{n \to m} = f_{n,r(1+s_d)}(m),
 \end{equation}
 Even if $s_d=0$ the total number of individuals after the decline fluctuates around the mean  $rN$. To compare our analytic expression with  Markov-matrix-based  numerics that requires exactly $rN$ individuals in the poor state we implemented slightly different decline procedure that have the same mean and variance, while in Monte-Carlo simulations we used the two binomial deviates. Details are given  in  Appendix \ref{S3}.

 Overall, the transition probabilities in each elementary time-step are,

\begin{eqnarray} \label{eq112233}
W^\pm_{n \to n} &=& (1-J^\pm)[1-2x(1-x)]  \nonumber \\
W^\pm_{n \to n+1}&=&(1-J^\pm) 2x(1-x) \left(\frac{1}{2}+\frac{s}{4}\right) \nonumber \\
W^\pm_{n \to n-1}&=&(1-J^\pm)2x(1-x)\left(\frac{1}{2}-\frac{s}{4}\right) \\
Q^-_{n \to m} &=& J^+ P^{decline}_{n \to m} \nonumber \\
\qquad Q^+_{n \to m} &=& J^- P^{growth}_{n \to m}. \nonumber
\end{eqnarray}

In these expressions $W^\pm_{n \to n_1}$ is the chance to reach $n_1$ from $n$ in an elementary duel when the environment is poor (minus, carrying capacity $rN$) or rich (plus, carrying capacity $N$). Correspondingly, $Q^{\pm}_{n \to  m}$ is the chance to reach $ m$ if the system was at $n$ right before a sharp decline  or a sharp increase.  

If at $t=0$ the mutant strain is represented by $n$ individuals, and the environment is in its rich/poor (plus/minus) state, the chance of fixation by the mutant, $\Pi^{\pm}_n$, satisfies the discrete  Backward Kolmogorov equation (BKE),
\begin{equation} \label{eq2}
\Pi^\pm_n = W^\pm_{n \to n+1} \Pi^\pm_{n+1}+ W^\pm_{n \to n-1} \Pi^\pm_{n-1}+ W^\pm_{n \to n} \Pi^\pm_{n} +  Q^\mp_{n \to m} \Pi^\mp_{ m}.
\end{equation}
This BKE is a $(N+rN-2) \times (N+rN-2)$ linear system  and may be solved numerically to obtain $\Pi^\pm_n$. This matrix-based technique was used to obtain the numerical results of Figs. \ref{fig1} and \ref{fig2} below, see Appendix \ref{S3}.

To solve the problem analytically, one would like to map the difference  equation (\ref{eq2}) into a differential equation. We define a \emph{significant step} as an interspecific duel (note that $n$ must change in such a significant step). For a given $x$, the chance of an elementary step (duel) to be  significant is $2x(1-x)$. The mean number of elementary steps needed for \emph{two} sharp  jumps (an increase and the following decrease or  vice versa) is $(N+rN)\tau = N(1+r)\tau$.
Therefore, the number of significant steps per a single decline/increase jump is,
\begin{equation} \label{eta}
\frac{1}{\eta}  \equiv x(1-x)N \tau (1+r).
\end{equation}
The chance of a single event to be a jump is therefore $\eta/(1+\eta)$ and its chance to be a competition step (duel) is $1/(1+\eta)$.   Moreover, for each significant step (interspecific duel) in the rich environment there are only $r$ significant steps in the poor environment.

Now we can write the  backward Kolmogorov equation for $\Pi(x) \equiv [\Pi^+(x)+\Pi^-(x)]/2$, i.e., for the chance to reach fixation when the initial state is in the plus or in the minus state with probability $1/2$.  When written per significant step it takes the form,
\begin{eqnarray} \label{eq5}
\Pi(x) &=& \frac{1}{1+\eta} \left[A \left(\frac{1}{2}+\frac{s}{4}\right) \Pi(x+1/N) + A\left(\frac{1}{2}-\frac{s}{4}\right) \Pi(x-1/N) \right.   \\ &+& \left. B \left(\frac{1}{2}+\frac{s}{4}\right) \Pi(x+1/rN) + B\left(\frac{1}{2}-\frac{s}{4}\right) \Pi(x-1/rN) \right] \nonumber  \\  &+& \frac{\eta}{2(1+\eta)} \left[\Pi \left(\frac{n+f_{N(1-r),x+s_g x(1-x)}(m)}{N}\right)+\Pi \left(\frac{f_{n,r(1+s_d)}(m)}{f_{n,r(1+s_d)}(m)+f_{N-n,r}(m')}\right) \right], \nonumber
\end{eqnarray}
where $A = 1/(1+r)$ and $B=r/(1+r)$. Note that in the sharp decline term  we used  two binomial deviates.

Next one would like to replace the binomial mass distribution in the last two terms by jumps to two possible destinations, one is the mean plus the standard deviation and the other is the mean minus the standard deviation, 
\begin{gather} \label{eq6}
\Pi \left(\frac{n+f_{N(1-r),x+s_g x(1-x)}(m)}{N}\right) \approx  \\ \nonumber  \frac{1}{2} \left[\Pi\left(x+s_g x(1-x)(1-r)+\sqrt{\frac{x(1-x)(1-r)}{N}}\right)+\Pi\left(x+s_g x(1-x)(1-r)-\sqrt{\frac{x(1-x)(1-r)}{N}}\right)\right] \\ \nonumber \Pi \left(\frac{f_{n,r(1+s_d)}(m)}{f_{n,r(1+s_d)}(m)+f_{N-n,r}(m')}\right) \approx \\ \frac{1}{2} \left[\Pi\left(x+s_d x(1-x)+\sqrt{\frac{x(1-x)(1-r)}{Nr}}\right)+\Pi\left(x+s_d x(1-x)-\sqrt{\frac{x(1-x)(1-r)}{Nr}}\right)\right] \nonumber
\end{gather}
Plugging Eq. (\ref{eq6}) and the value of $\eta$ as defined in Eq. (\ref{eta})  into (\ref{eq5}), and expanding all functions to second order in $1/N$ and to first order in selection terms yields,
\begin{equation} \label{eq7}
\left(\frac{1}{2r}  + \frac{1-r}{4 \tau r}  \right) \frac{\Pi''(x)}{N^2}  + \left(\frac{s}{1+r} + \frac{s_g(1-r)+s_d}{2  \tau (1-r) } \right) \frac{\Pi'(x)}{N} = 0.
\end{equation}
In Eq. (\ref{eq7}) we assumed that jumps are relatively rare and $\eta \ll 1$ so $1-\eta \approx 1$, otherwise the effect of equilibrium competition is negligible in comparison to the effect of the jumps. Note that the condition $s_d<(1-r)/r$ ensures that in the limit $r=1$ the contribution from $s_d$ vanishes.

Similar considerations for $T_A(x) \equiv [T_A^+(x)+T_A^-(x)]/2$ yield,
\begin{equation} \label{eq7t}
\left(\frac{1}{2r}  + \frac{1-r}{4 \tau r}  \right) \frac{T_A''(x)}{N^2}  + \left(\frac{s}{1+r} + \frac{s_g(1-r)+s_d}{2  \tau (1+r) } \right) \frac{T_A'(x)}{N} = -\frac{1}{N(1+r)x(1-x)}.
\end{equation}
The last term in Eq. (\ref{eq7t}) reflects the mean time for a significant step, which is $1/2Nx(1-x)$ with probability $1/(1+r)$  and $1/2rNx(1-x)$ w.p. $r/(1+r)$.

 Comparing Eq. (\ref{eq7}) with Eq. (\ref{eq0}) and equation (\ref{eq7t}) with Eq. (\ref{eq0t}),  one finds that the solution still has the form of the solution in static environment.
  Accordingly, one can implement the \emph{static} environment formulae (like Eq. \ref{eq2new} above)  to obtain the $\Pi(x)$, $T_A(x)$ and $T_F(x)$,  by replacing the value of $N$ and $s$  by their effective counterparts,
\begin{equation} \label{eq1new}
N_{eff} = N \frac{4r \tau}{(1+r)(2 \tau +1-r)}.
\end{equation}
and
\begin{equation} \label{eq1newsel}
s_{eff} = s+\frac{s_g(1-r)}{2\tau}+\frac{s_d}{2\tau}.
\end{equation}

\begin{figure*}[t]
	\centering{
		\includegraphics[height=4.3cm,width=\textwidth]{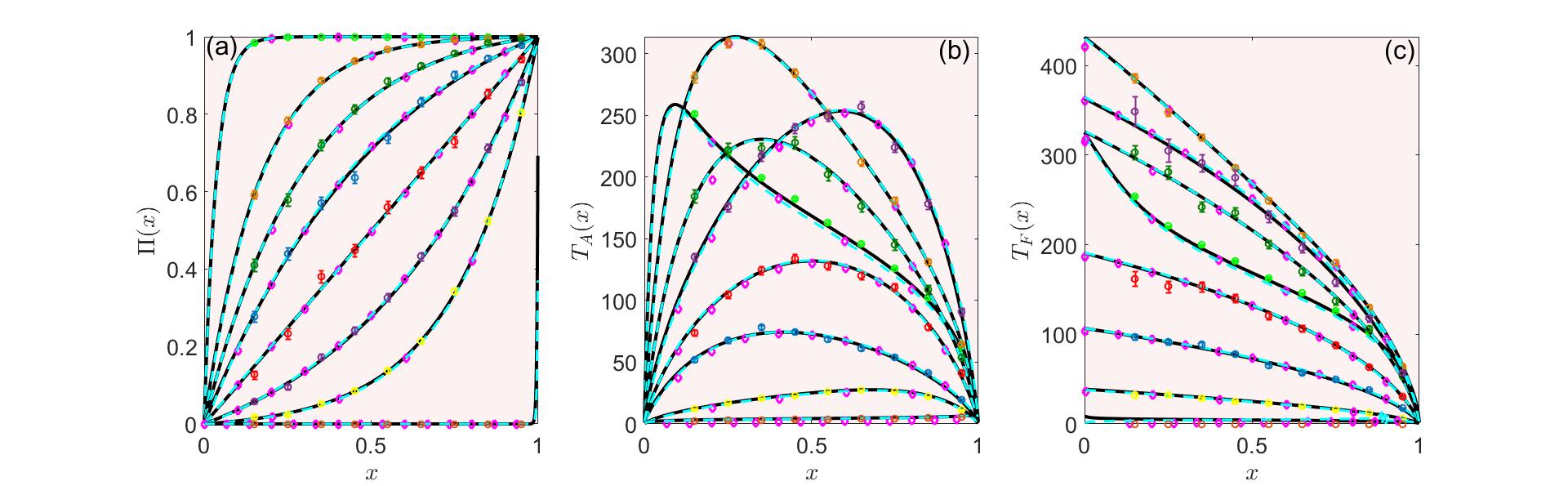} \\
\includegraphics[height=4.7cm,width=\textwidth]{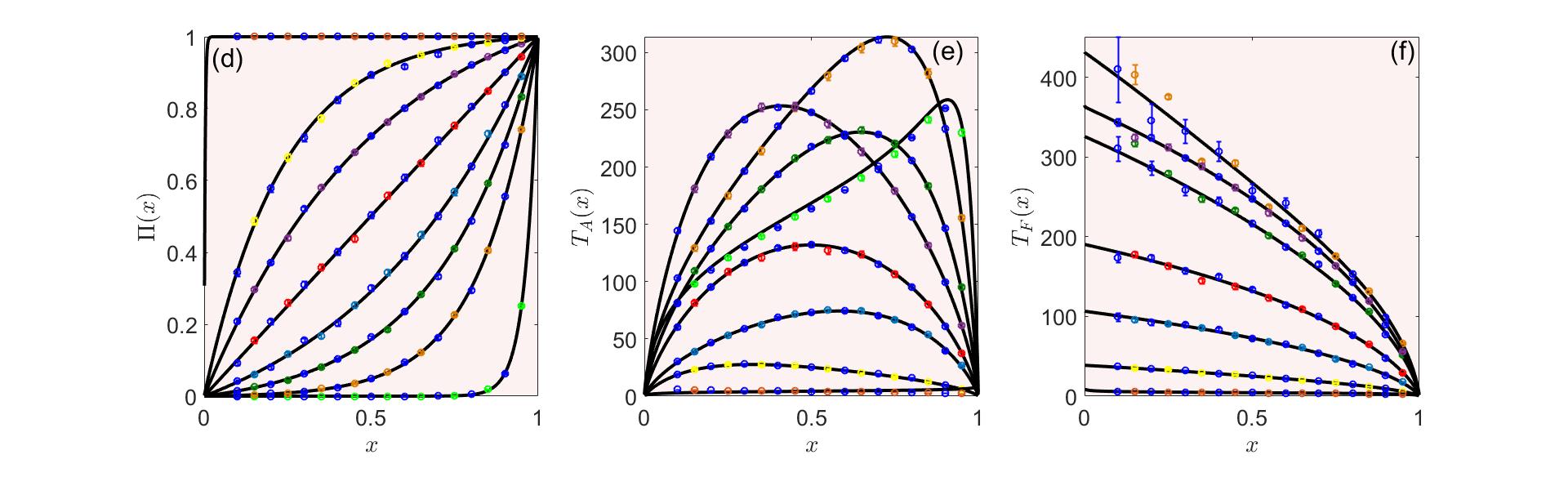}}
	\caption{$\Pi(x)$ (panels a and d) $T_A(x)$ (b and e) and $T_F(x)$ (c and f)  as a function of $x$ for various values of $N$, $\tau $, $r$ and selection coefficients. Results for stochastic environment are presented in panels a-c, while the results for periodic variations are shown in panels d-f.  For each set of parameters, the solid black line represents the relevant analytic prediction [Eq. (\ref{eq2new}) for $\Pi$, (\ref{eq3new}) for $T_A$  and (\ref{eq4new}) for $T_F$, where $N$ is replaced by $N_{eff}$ from Eq. (\ref{eq1new}) and $s$ is replaced by $s_{eff}$ from (\ref{eq1newsel})].  In panels a-c  (stochastic environment) we show  numerical solutions of a Markov matrix model with local (dashed cyan curves) and global (magenta diamonds) competition, together with MC simulation results for local competition (colored circles, plotted with one std error bars that are  usually too small to be seen). In panels d-f (periodic environment), analytic results are compared with MC simulations for local (colored circles) and global (blue circles) competition. Our numerical procedures are explained in Appendix \ref{S3}. Parameters ranges are $N \in [1000,4000]$, $r \in [0.2,0.75]$,  and $\tau \in [0.05,5]$, the  details for each curve are listed in Appendix \ref{S2}. Parameter set 1 (see Appendix \ref{S2}) is missing in panel (f), since the chance of fixation is too small.   The red lines were obtained for combinations of $s$, $s_d$ and $s_g$ that yields $s_{eff}=0$.       \label{fig1}}
\end{figure*}

\section{The diffusion approximation and its range of applicability} \label{sec4}

 Eqs. (\ref{eq1new}) and (\ref{eq1newsel}) are our main results. Their derivation is based on two  assumptions
  \begin{itemize}
    \item We assumed that the number of up-down flips before fixation is large, i.e., that the system is in its ``micro-evolutionary" (annealed) regime as defined in \cite{mustonen2008molecular}. When the chance of fixation to take place during a single sweep is large, $\Pi$ depends not only on $x$ but also on the initial state of the environment.
    \item We assumed that the population is large (both $N$ and $rN$ are much larger than one) and the selection coefficients are small, $s, s_g,s_d \ll 1$.
  \end{itemize}
   Importantly, when this procedure is adequate  the assumptions of local competition and stochastic fluctuations are unnecessary. The diffusion approximation implies that the only important factors are the mean  and the variance of $\Delta x$ (the change in $x$ per event), and since these different models yield the same mean and the same variance (to the leading order in $s$, $s_d$, $s_g$ and $1/N$), the outcome is independent of the details of the dynamics.

This amazing property is demonstrated in panels (a) and (d) of Fig. \ref{fig1}. The numerical results (obtained by inverting a Markov matrix or from direct Monte-Carlo simulations, as explained in Appendix \ref{S3}) show perfect agreement with the analytic predictions for deleterious, beneficial and neutral mutants, under global and local competition and  in periodic or stochastic environment.

Similarly, one may obtain the mean time to absorption $T_A(x)$ and  the mean time to fixation $T_F(x)$  using the static environment formulae with $N_{eff}$ and $s_{eff}$ as defined in Eqs. (\ref{eq1new}) and (\ref{eq1newsel}). Expressions for  $T_A(x)$ and $T_F(x)$ in static environment were presented in \cite{crow1970introduction} and in Appendix \ref{S4} we derive  simpler expressions for these quantities, using the exponential integral $E_i(x)= -\int_{-x}^\infty  dt \ \exp(-t)/t$. These are,

\begin{widetext}
\begin{equation} \label{eq3new}
T_A(x) = C_2 + C_1 e^{-sNx} + \frac{1}{s} \left(e^{-sNx} E_i(sNx) - e^{sN(1-x)} E_i[-sN(1-x)] +\ln \left[\frac{ 1-x}{sNx} \right] \right),
\end{equation}
and,
\begin{eqnarray}\label{eq4new}
T_F &=& \frac{C_4 + C_3 e^{-sNx} + \frac{1}{s} \left(e^{-sNx} E_i(sNx) - e^{sN(1-x)} E_i[-sN(1-x)] +\ln \left[\frac{ 1-x}{sNx} \right] \right)}{\Pi(x)} \nonumber \\ &+& \frac{E_i[-sNx]-e^{-sN}Ei[sN(1-x)] - e^{-sNx}\ln[x/(1-x)]}{sN\Pi(x)}.
\end{eqnarray}
\end{widetext}
The constants $C_1 ... C_4$ are given in Appendix \ref{S4}. Again, panels (b-c, e-f) of Figure \ref{fig1} demonstrate the agreement between these formulae (with $N \to N_{eff}$ and $s \to s_{eff}$) and the numerical result for $T_A$ and $T_F$ in different scenarios.

\begin{figure}[h]
	\centering{
		\includegraphics[width=7cm]{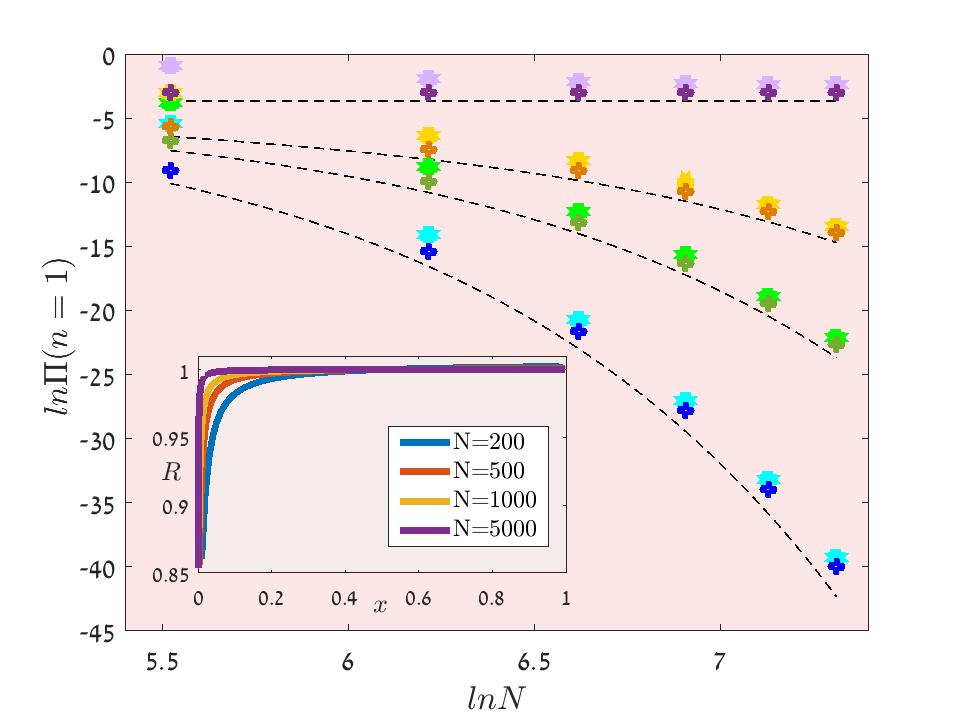}}
	\caption{In the inset we present $R$ vs. $x$, where $R$ is the ratio between our analytic expression for $\Pi(x)$ [Eq. (\ref{eq2new}) with the $N_{eff}$ given in (\ref{eq1new})] and its numerically calculated value for stochastic environment obtained from inversion of the Markov matrix (see Appendix \ref{S3}).  Parameters are  $r=\tau = 0.5$, $s_g=s_d=0$.  Different values of $N$ were used (see lagend), where the value of $sN$ is kept constant ($sN=-0.1$), so the analytic prediction is $N$ independent. Clearly, the ratio becomes closer and closer to one as $N$ increases, meaning that the quality of our approximation improves with $N$. The results for a single mutant ($n=1$ or $x=1/N$) show the slowest convergence. In the main figure this ``worst case scenario" is considered. The logarithm of $\Pi(n=1)$ is plotted against $\ln N$, and the analytic prediction (dashed black lines) is compared with the numerical results for the local (light circle) and global (dark crosses) competition models. All the results were obtained for $\tau = 0.1$ and $r=0.5$, $s_g=s_d=0$. Markers with different colors stand for   $s=-0.2$ (blue). $s=-0.1$ (green), $s=-0.05$ (brown) and $s=0.2$ (purple).       \label{fig2}}
\end{figure}

Since our expressions  for $N_{eff}$ and $s_{eff}$ were derived using the continuum (diffusion) approximation, the agreement between the analytic formula and the numerical results becomes better as $N$ grows, as demonstrated in Figures \ref{fig2}.

The equivalence of stochastic and periodic variations (when the parameters of the process are calibrated appropriately, as done here) appears to be a generic feature of the approximation used.  When the diffusion approximation holds, each ``elementary"  event (rapid growth, rapid decline, equilibrium dynamics over $\tau$ generations) contributes a given amount of mean change in $x$ to the coefficient of $\Pi'(x)$ (or $T_A'$ etc.), and given variance to the coefficient of $\Pi''(x)$. Since we kept only terms that are linear in $s$ (or $s_d$ or $s_g$) and in $1/N$, each elementary event may split into two or more sub-events without changing the corresponding BKE. When $s^2 $ contributions are important this is not the case anymore, as equilibrium competition for time $\tau$ and two $\tau/2$ period contribute differently to $s^2$ terms.  Nonlinear effects  are important when equilibrium selection changes sign~\cite{yi2013bounded,danino2018fixation,meyer2018noise} and control the efficiency of bet-hedging strategies like phenotypic switching~\cite{thattai2004stochastic,patra2015emergence}.

\section{Discussion}

Our system admits two qualitative types of demographic processes: population jumps (sharp growth or sharp decline) and equilibrium steps, associates with birth-death competition events, between jumps. When $\tau$ is very small $N_{eff} \sim N \tau$ decreases substantially together with the values of $T_A$ and $T_F$ that scale like  $N_{eff}$ (for weak selection) or $\ln(N_{eff})/s_{eff}$ (for strong selection).  For some biological species the effective population size (as inferred from genetic polymorphism data) is much smaller than the adult (census) population size~\cite{hauser2008paradigm}. Our model suggests that this phenomenon may be related to fast jumps associated with rapidly fluctuating environmental conditions.  Fast jumps also wipe out the effect of $s$, so $s_{eff}$ is determined only by $s_g$ and $s_d$. 

In the opposite limit, when $\tau$ is very large, the jump processes (both the jump noise and the selection during the jumps as expressed in $s_d$ and $s_g$) becomes negligible. In that case $N_{eff}$ is the harmonic mean of $N$ and $rN$, while $s_{eff} = s$.

The probability of ultimate fixation for a  deleterious mutant in fluctuating environment was considered recently by Wienand and coworkers~\cite{wienand2017evolution,wienand2018eco} (their ``pure resource competition scenario").  These authors studied a  Moran process of two strains with constant, fitness dependent, birth rate ($b=1$ for the wild type and $b=1+s$ for the mutant species) and  temporally varying, fitness independent death rate. In their work,  $\Pi$ was approximated by the integral $\int  \ dN \ \Pi(N) P_{s \tau}(N)$, where $\Pi(N)$  is the static environment formula, Eq. (\ref{eq2new}) and $P_{s \tau}(N)$ is the probability to find the system with $N$ individuals when the mean time between two environmental switches is $s \tau$. The relationships of this approximation and our technique require additional examination.

We would like to stress that the two extreme limits (large and small $\tau$) of the model of Wienand and coworkers are not the limiting cases considered above.  In \cite{wienand2017evolution,wienand2018eco} $\tau_{ad}$, the adaptation time required for the system to adapt to  the new carrying capacity ($N \rightleftarrows rN$) after an environmental shift, is about one generation. Accordingly, when $\tau$ approaches zero $P_{s \tau}(N)$ becomes a delta function (or almost a delta function, due to demographic stochasticity) at the harmonic mean of the carrying capacities. In our calculations $\tau_{ad} \to 0$ (sharp growth or decline), so population jumps occur for any $\tau$.   Similarly, in the case $\tau \to \infty$ considered in \cite{wienand2017evolution,wienand2018eco} fixation takes place during a single sweep of the environment (this is the macro-evolutionary case considered in~\cite{mustonen2008molecular}  and \cite{marrec2019resist}). The implementation of the diffusion approximation in our model requires that, although $\tau$ is large with respect to other constants, it still must be much smaller than $T_F$ to allow for many jumps before fixation, hence (unlike \cite{wienand2017evolution,wienand2018eco}) we did not obtain the arithmetic mean ($\Pi = [\Pi(N)+\Pi(rN)]/2$) in that limit. Incidently, we obtained the $\tau \to 0$ outcome of \cite{wienand2017evolution,wienand2018eco}  ($N_{eff}$ is the harmonic mean) in the opposite limit where $\tau$ is very large.

To better account for  realistic situations, a few extensions of our model are required. First, we considered here only instantaneous growth (and decline) events. In reality growth of a population takes time, and abundance fluctuations that occur early during the growth period are amplified, leading to larger variance per growth event. Second, we assumed that the rate of events at equilibrium is proportional to the size of the population, so there are more elementary birth-death events per unit time in the rich state. In some realistic scenarios (e.g., when death happens due to predation and the number of predation events is independent of the population size) this may not be the case. Third, the equilibrium selection may depend on time. $s$ may have different magnitude and/or sign in the rich and in the poor state, or its value may fluctuate regardless of the state of the  carrying capacity. 

Still, the tractability  of our models, and the robustness  of our formula against modifications of the underlying dynamics, reflects an inherent feature. In these systems, the only important parameter  is the relative strength of stochastic vs. deterministic factors.  This characteristics  is determined by the ratio between the number of selection steps and genetic drift steps. As a result, the solution was obtained  through a simple step counting argument. We believe that the method presented here is applicable to all the extensions mentioned above, and intend to address these extended models  in a subsequent publication.

\section{Acknowledgments}
  We thank Benjamin Good and David Kessler for  helpful discussions. This research  was supported by the ISF-NRF Singapore joint research program (grant number 2669/17).The work of I.M. was supported by an Eshkol Fellowship of the Israeli Ministry of Science.

\bibliography{refs}

\clearpage
\setcounter{figure}{0}
\setcounter{table}{0}
\renewcommand{\thetable}{A\arabic{table}}
\renewcommand{\thefigure}{A\arabic{figure}}
\appendix

\section{Detailed description of the parameters sets used in Fig. \ref{fig1} } \label{S2}

Figure \ref{fig1} of the main text is reproduced here with a number attached to each dataset. The parameters are given for each number in Table \ref{table1} below.

\begin{figure*}[h]
	\centering{
		\includegraphics[width=\textwidth]{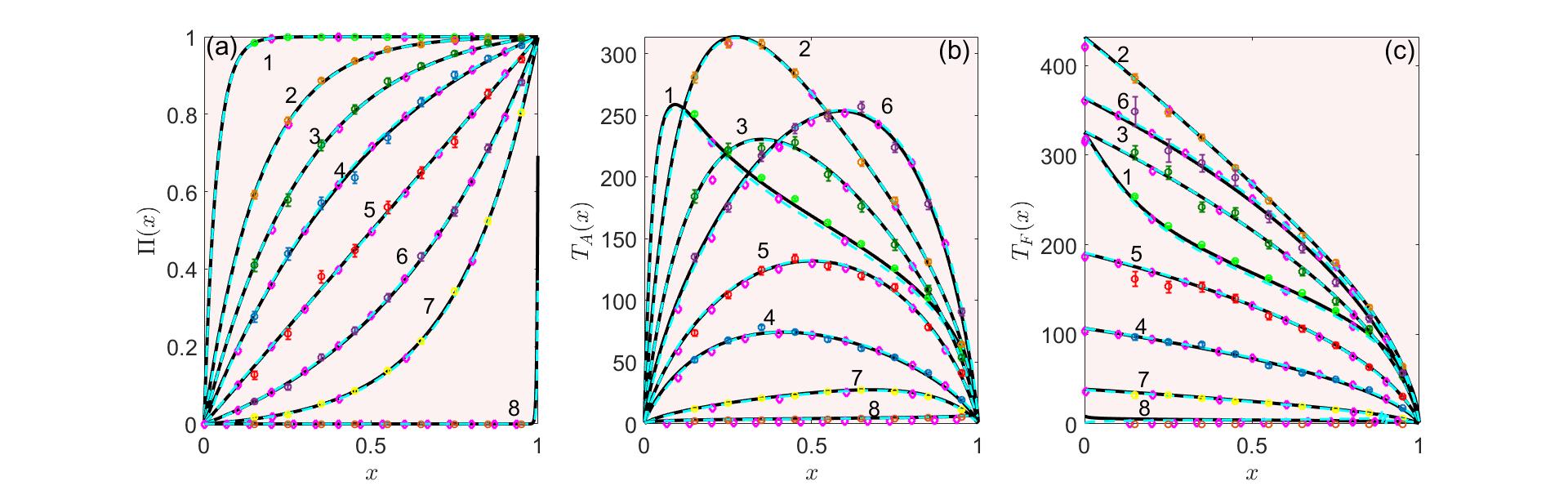} \\
        \includegraphics[width=\textwidth]{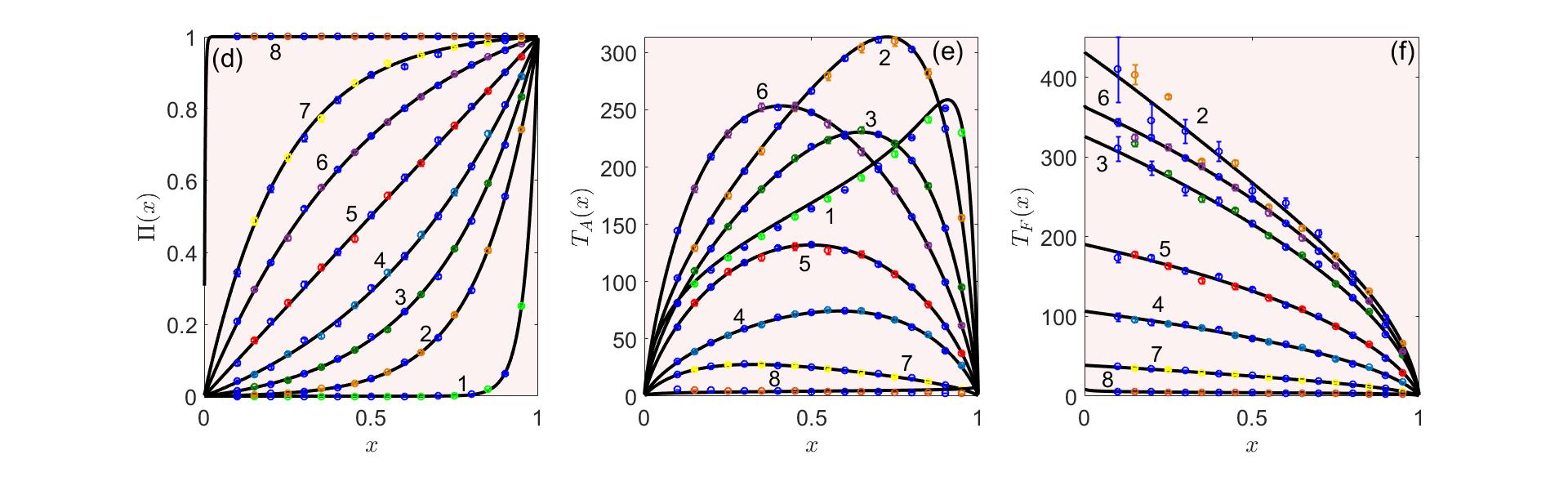}}
	\caption{$\Pi(x)$ (panels a and d), $T_A(x)$ (b and e) and $T_F(x)$ (c and f)  as a function of $x$ for various values of $N$. The number attached to each line in panels a-c correspond to a different set of parameters, detailed in table \ref{table1}. Panels d-f were obtained for exactly the same set of parameters, except that the direction of selection was reversed (i.e., $s \to -s$, $s_g \to -s_g$ and $s_d \to -s_d$).    \label{figS1}}
\end{figure*}

\begin{table}[h]
\begin{center}
\caption {Parameters: Fig. \ref{fig1}} \label{table1}
   \begin{tabular}{| >{\centering\arraybackslash}m{2cm} || >{\centering\arraybackslash}m{2cm} |>{\centering\arraybackslash}m{2cm} |>{\centering\arraybackslash}m{2cm} | >{\centering\arraybackslash}m{2cm} |>{\centering\arraybackslash}m{2cm} |>{\centering\arraybackslash}m{2cm} | >{\centering\arraybackslash}m{2cm} |}
    \hline
    index &  color &  $N$ & $\tau$	 & $r$ & $s$ & $s_g$ & $s_d$ \\ \thickhline
    1 & light green & 4000& 1&		 0.25& 	 0.01 & 			-0.03 & 0.05 \\ \hline
    2 & light brown & 2500& 1&		 0.2& 	 0.01 & 			0 & 0 \\ \hline
    3 & dark green &  1000& 5&	 	0.25& 	0.01 &			 	0.0 & -0.01 \\ \hline
    4 & turquoise &  1000& 0.05& 		0.5& 	 0.001 & 			0.001 & 0.001 \\ \hline
    5 & red & 1000& 0.1& 		0.5&	 	 -0.002 & 		0.01 &  -0.0046\\ \hline
    6 & purple & 2000& 0.1&		 0.5&	 	 -0.005 & 	0.000001 & -0.00001 \\ \hline
    7 & yellow  & 1000& 0.05& 		0.25&	 	 0.01 & 		0 & -0.01 \\ \hline
    8 & dark brown & 1000& 0.05& 		0.75&	 	 0 & 		0.2 & -0.2 \\ \hline

    \end{tabular}
\end{center}
\end{table}

\clearpage

\section{Numerical techniques} \label{S3}

Through this paper we compared the  analytic predictions [Eqs. (\ref{eq2new},\ref{eq3new},\ref{eq4new}) of the main text, with $N_{eff}$  and $s_{eff}$] with numerical results that were obtained from two types of numerical calculations. Here we provide some details for the numerics.

\subsection{Markov matrix inversion}

 The essence of this technique is explained in Eq. (\ref{eq2}) above and in the following text, for more details and examples  see Appendix A2 of \cite{danino2016stability} or Appendix A of \cite{yahalom2019comprehensive}.

A technical problem appeared when we tried to implement our method to the sharp decline period. If the number of mutant individuals is picked at random from $B_{xN}[r(1+s_d)]$ and the number of wild type is picked independently from $B_{(1-x)N}[r]$, the total size of the population after the decline fluctuates around $Nr$. To implement our exact numeric technique the total population size after the decline period must be \emph{exactly} $Nr$, so we cannot use two independent binomial trials.

To overcome this difficulty we assumed that in a decline step,
\begin{equation}
n \to r\frac{n^2}{N} + s_d r n(1-n/N)+ B_{\frac{n(N-n)}{N}}[r].
\end{equation}
so,
 \begin{equation}
 P^{decline}_{n \to m} = f_{Nx(1-x),r}(m-rNx^2 +s_drNx(1-x)).
 \end{equation}
 When $s_d <<1$ this process has the same mean and variance of the two independent binomial trial processes, and since the solution involves only the mean and the variance, it yields the same outcomes.

\subsection{Monte-Carlo simulations}

We have implemented direct MC simulations for two reasons. First, the Markov matrix inversion numerics is applicable only in the stochastic case, when per each step there is a finite probability that the system switches from poor to rich state or vice versa. To implement the same matrix technique in the periodic case one must calculate first the corresponding Floquet operator for equilibrium periods of duration $\tau$. Instead, we implemented direct MC simulations.

Secondly, the Markov matrix technique is applicable only if the system jumps between two states, one with exactly $N$ individuals and the other with $rN$. As we have seen above, this restriction is incompatible with the realistic modelling of a sharp decline periods, where the chance of each individual to survive is $r$, so we implemented an ad-hoc binomial mass function which has the same mean and variance. In the MC simulation, on the other hand, decline involves two independent binomial deviates, one for the mutant and one for the wild type, without any global restriction.

\clearpage

\section{Derivation of $T_A$ and $T_F$ in static environment} \label{S4}

The formulae derived below are our simplified version of the solutions presented in \cite{crow1970introduction}, page 430.

\subsection{Calculation of $T_A$}

To calculate $T_A$ in static environmental conditions ($s$ and $N$ are time independent) one writes the Backward Kolmogorov Equation (BKE). Starting with  $n$ individuals, the time to either fixation or extinction is $T_n = \sum W_{n,m} T_m + \Delta t$, where $\Delta t$ is the time needed for a step (in our case, $\Delta t = 1/N$). The sum is over all possible destination states (in our case, $m \in [n-1,n,n+1]$  and the $W$s are the transition probabilities defined in Eq. \ref{eq112233}. Accordingly,
\begin{equation}
T_n = 2x(1-x) \left(\frac{1+s/2}{2}T_{n+1} + \frac{1-s/2}{2}T_{n-1} \right) + [1-2x(1-x)]T_n +\frac{1}{N}.
\end{equation}
In the continuum limit, $x=n/N$ and  $T_{n \pm 1} \approx T(x) \pm T'(x)/N + T''(x)/(2N^2)$,
\begin{equation} \label{same}
T'' + sNT' = -\frac{N}{x(1-x)}.
\end{equation}
The boundary conditions are of course $T(0) = T(1) = 0$.

Using the integration factor $\exp(sNx)$ one may find $T'$
\begin{equation}
T' = C e^{-sNx} -N e^{-sNx} E_i(sNx)+  N e^{sN(1-x)} E_i(-sN[1-x]),
\end{equation}
where we used the exponential integral
\begin{equation}
E_i(x)= -\int_{-x}^\infty  dt \frac{e^{-t}}{t}.
\end{equation}

Second integration and implementation of the boundary conditions yields Eq. (\ref{eq3new}) of the main text with,
\begin{eqnarray}
C_1 &=& \frac{-2 e^{sN} \gamma_E + e^{2sN} E_i(-sN) + E_i(sN) - 2 e^{sN} \ln(sN)}{s(e^{sN}-1)} \nonumber \\
C_2 &=& \frac{\gamma_E (1+e^{sN})-e^{sN}[2 \ln(sN)-E_i(-sN)] -  E_i(sN)}{s(e^{sN}-1)}.
\end{eqnarray}

\subsection{Calculation of $T_F$}

To calculate $T_F$ one defines~\cite{redner2001guide}  $Q = \Pi \cdot T_F$. $Q$ satisfies,
\begin{equation}
Q'' + sNQ' = -\frac{N }{x(1-x)}+\frac{N e^{-sNx} }{x(1-x)},
\end{equation}
with the boundary conditions $Q(0)=Q(1)=0$. Except of the last term, we obtained the same equation  as Eq. (\ref{same}), meaning that
\begin{equation}
Q'(x) = T_A'(x) + N \ln\left(\frac{x}{1-x}\right) e^{-sNx}.
\end{equation}
Accordingly, one obtains Eq. (\ref{eq4new}) of the main text with

\begin{eqnarray}
C_3 &=& \frac{-3 \gamma_E  e^{sN }-3 e^{sN } \ln (sN )-\ln (sN )+e^{sN} \text{Ei}(-sN)+e^{2 sN } \text{Ei}(-sN )+2 \text{Ei}(sN )-\gamma_E }{\left(e^{sN }-1\right) s}  \\
C_4 &=& \frac{e^{-sN } \left(3 \gamma_E  e^{sN }+\gamma_E  e^{2sN }+2 e^{sN } \ln (sN )+2 e^{2 sN } \ln (sN )-2 e^{2 sN} \text{Ei}(-sN)-e^{sN } \text{Ei}(sN )-\text{Ei}(sN )\right)}{\left(e^{sN }-1\right) s} \nonumber
\end{eqnarray}

\end{document}